\documentclass[aps,twocolumn,amsmath,amssymb,nofootinbib,superscriptaddress,prl]{revtex4-2}
\usepackage[colorlinks,
            linkcolor=black,
            citecolor=black,
            urlcolor=black
            ]{hyperref}
\usepackage{subfigure}
\usepackage{amsthm,amsmath,amssymb}
\usepackage{mathrsfs}
\usepackage{braket}
\usepackage{ulem}
\usepackage{graphicx}
\usepackage{dcolumn}
\usepackage{bm}
\usepackage[noend]{algpseudocode}
\usepackage[ruled,vlined]{algorithm2e}
\usepackage{multirow}
\usepackage{color}
\usepackage{booktabs}
\usepackage{longtable}
\usepackage{hyperref}
\definecolor{Red}{rgb}{1,0,0}

\begin{document}
\title{Quantum Flow Matching}

\author{Zidong Cui}
\affiliation{Institute of Fundamental and Frontier Sciences, University of Electronic Science and Technology of China, Chengdu 611731, China}
\affiliation{School of Physics, University of Electronic Science and Technology of China, Chengdu 611731, China}

\author{Pan Zhang}
\affiliation{Institute of Theoretical Physics, Chinese Academy of Sciences, Beijing 100190, China}
\affiliation{School of Fundamental Physics and Mathematical Sciences, Hangzhou Institute for Advanced Study, UCAS, Hangzhou 310024, China}

\author{Ying Tang}

\email[Corresponding authors: ]{jamestang23@gmail.com}
\affiliation{Institute of Fundamental and Frontier Sciences, University of Electronic Science and Technology of China, Chengdu 611731, China}
\affiliation{School of Physics, University of Electronic Science and Technology of China, Chengdu 611731, China}
\affiliation{Key Laboratory of Quantum Physics and Photonic Quantum Information, Ministry of Education, University of Electronic Science and Technology of China, Chengdu 611731, China}
\affiliation{Non-classical Information Science Basic Discipline Research Center of Sichuan Province, University of Electronic Science and Technology of China, Chengdu 611731, China}

\begin{abstract}
The flow matching has rapidly become a dominant paradigm in classical generative modeling, offering an efficient way to interpolate between two complex distributions. We extend this idea to the quantum realm and introduce the Quantum Flow Matching (QFM), a quantum-circuit realization that offers efficient interpolation between two density matrices. QFM offers systematic preparation of density matrices and generation of samples for accurately estimating observables, and can be realized on quantum computers without the need for costly circuit redesigns. We validate its versatility on a set of applications: (i) generating target states with prescribed magnetization and entanglement entropy, (ii) estimating nonequilibrium free-energy differences to test the quantum Jarzynski equality, and (iii) expediting the study on superdiffusion. These results position QFM as a unifying and promising framework for generative modeling across quantum systems.
\end{abstract}

\maketitle

\section{I. Introduction}
The density matrix $\rho$  is key to describing finite-temperature quantum systems~\cite{RevModPhys.83.863,D'Alessio03052016}.  Preparing a target $\rho=\sum_{i}p_{i}\ket{\psi_{i}}\bra{\psi_{i}}$ as an ensemble of pure states and estimating observables from it~\cite{banuls2020simulating} depend on two critical factors: determining the probability $p_{i}$ and preparing the pure state $\ket{\psi_{i}}$. Modern generative models~\cite{RevModPhys.91.045002} offer a route to capture both factors, including classical neural networks for ground-state search~\cite{PhysRevLett.121.167204,torlai2018neural} and tracking $\rho$'s dynamics~\cite{schmitt2025simulating}; tensor networks for thermal state simulation~\cite{PhysRevLett.93.207204}; and quantum-classical hybrid models such as $\beta$-VQE~\cite{liu2021solving} that integrate quantum circuits to estimate observables. Yet, these approaches are largely classical, leaving fully quantum protocols less explored. Typical quantum methods approximate $p_{i}$ via statistical sampling, such as the minimally entangled typical thermal state (METTS) algorithm~\cite{stoudenmire2010minimally,PhysRevLett.102.190601} that employs a Markov chain. However, preparing states in the ensemble remains nontrivial and costly, since each pure state demands a dedicated quantum circuit, leading to repeated circuit adjustments and significant overhead, especially in tasks such as estimating free energy changes~\cite{Qu_Jar_exp_1} and studying phase transitions~\cite{getelina2023adaptive}. These challenges call for more efficient fully-quantum protocols to prepare and track $\rho$.

A quantum generative model that prepares $\rho$ with a single circuit would be an efficient framework aligned well with quantum platforms, potentially bypassing the bottleneck of circuit adjustments. The recent quantum denoising diffusion probabilistic model (QuDDPM)~\cite{QuDDPM} generates state ensembles by evolving from Haar-random states~\cite{brandao2016local,cwiklinski2013local}, analogous to Gaussian noise classically. However, QuDDPM’s inverse sampling prepares only the target ensemble, without capturing the time evolution of the target system, limiting its use in scenarios requiring flexible initialization or precise state control~\cite{Qu_sim_1,schuckert2025observation,kumaran2025quantum}. In contrast, the classical flow matching~\cite{ma2024sit,Flow_matching_GM,Vari_infer_Nor_Flow}, which learns flows to map between distributions, provides an effective alternative. Extending this idea to the quantum realm may overcome these limitations and broaden applications in quantum dynamics.

In this work, we develop quantum flow matching (QFM). While classical flow matching learns a flow that transports probability distributions, QFM learns a density-matrix propagator implemented by a quantum circuit, under constraints of circuit depth, measurements, and sampling. The propagator can be obtained in a data-driven manner or analytically, depending on the task. As classical flow matching generalizes diffusion models, QFM generalizes QuDDPM by including it as the special case starting from a Haar-random ensemble, and enables the time evolution from any initial ensembles toward target ones. We benchmark QFM against QuDDPM on the topological-state evolution, the entanglement growth, and the transverse-field Ising model (TFIM) (Appendix), and then apply it to more challenging tasks, including the thermal-state preparation for free-energy estimation~\cite{Qu_Jar_exp_1} and the superdiffusive dynamics~\cite{kumaran2025quantum}, both implemented with a fixed circuit. We further provide scaling analysis and bounds on the generalization error of QFM, supporting its wider applicability. For clarity, all mathematical symbols used in this work are summarized in Table~\ref{tab:notion}.

\begin{figure*}[!htp]
    \centering
    \includegraphics[width=1\linewidth]{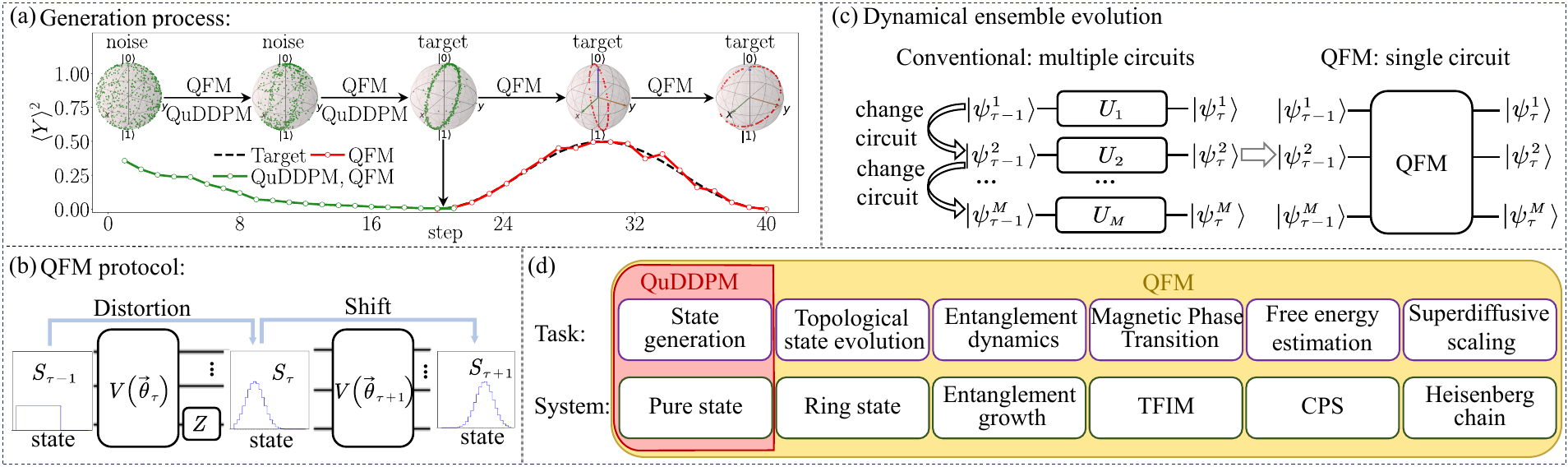}
    \caption{Schematic of the quantum flow matching. (a)Inspired by classical flow matching~\cite{Flow_matching_GM}, QFM learns the evolution between two density matrices. Unlike QuDDPM~\cite{QuDDPM}, it does not require starting from a Haar-random ensemble, allowing wider applications. (b) QFM alternates partly measured and unitary circuits to ensure convergence and stability. (c) A single circuit emulates density-matrix dynamics, reducing circuit adjustments; here $i$ labels the time steps and $M$ denotes the ensemble size. (d) Applications include free-energy estimation and superdiffusive scaling (main text), as well as topological-state evolution, entanglement growth, and magnetic phase transitions (Appendix).}
        \label{fig:QFM}
\end{figure*}

\
\begin{longtable}{cc}
\caption{Glossary of mathematical terms.} \label{tab:notion} \\ 
\toprule
Notion & Description \\ 
\midrule
\endfirsthead

\multicolumn{2}{c}%
{{ \tablename\ \thetable{} -- continued from previous page}} \\
\toprule
Notion & Description \\ 
\midrule
\endhead
\midrule \multicolumn{2}{r}{{Continued on next page}} \\ 
\endfoot
\bottomrule
\endlastfoot
$\rho_{0}$ & the density matrix at step $0$ \\
$n$ & the number of data qubits \\
$n_a$ & the number of ancilla qubits \\
$\sigma_{z}^{i}$ & Pauli operator $Z$ on qubit $i$ \\
$S_{\tau}$ & the state ensemble at step $\tau$ \\
$M$ & number of states in one ensemble \\
$\ket{\psi_{\tau}^{m}}$ & a sample state in $S_{\tau}$ \\ 
$\mathcal{T}$ & the number of steps \\
$\theta$ & one parameter in $\vec{\theta}_{\tau}$ \\
$\vec{\theta}_{\tau}$ & the parameter vector of step $\tau$ \\
$\vec{\theta}_{\tau}^{op}$ & the optimized parameter vector \\
$V(\vec{\theta}_{\tau})$ & the circuit in step $\tau$ of QFM \\
$L$ & the layer number of EHA in each step \\
$U_{n_a}(\vec{\theta}_{\tau})$ & the circuit with ancilla qubits \\
$U_{n}(\vec{\theta}_{\tau})$ & the circuit only acts on data qubits \\ 
$U^{(1)}$ & single-qubit gate in EHA \\
$U^{(2)}$ & two-qubit gate in EHA \\
$U_{(r)}$ & the measurement-induced gate \\
$r_{n_a}$ & the measurement result of the $n_a$-th qubit \\
$R_{y}$ & the rotation operator about the $y$ axis \\ 
$\ket{\widetilde{\psi}_{\tau}^{m}}$ & the generated state in step $\tau$ \\
$\widetilde{S}_{\tau}$ & the generated ensemble in step $\tau$ \\
$M_{z}$ & measurement along the $Z$ direction \\
$D(\vec{\theta}_{\tau})$ & loss function \\
$H(t)$ & time-dependent Hamiltonian \\
$g(t)$ & transverse field strength \\
$\beta_{\tau}$ & inverse temperature \\
$J$ & 1D-interaction strength \\
$J_{\perp}$ & 2D-interaction strength \\
$K$ & number of different interaction ratio $J_{\perp}/J$ \\
$C_{22}^{J_{\perp}/J}$ & correlation function with prob qubit 2 \\
$a,b,c$ & the bonds of the Hamiltonian \\
\end{longtable}

\begin{table*}[hbtp!]
    \centering
    \caption{Hyperparameters and results in three benchmark examples. Here, $n$ is the number of data qubits, $n_a$ is the number of ancilla qubits,  $M$ is the number of training states, and $L$ is the number of layers in each step of QFM.}
    \label{tab:comparison}
    \begin{tabular}{ccccccccc}
        \toprule
        Task & $n$ & $n_a$   &$L$&$O$&$f$& $\mathcal{T}$& $M$ &Performace\\
        \midrule
        Ring state ensemble evolution & $1$& $0$&$5$& Swap test&Arithmetic Mean& 20& 100   &0.037\\
        Entanglement growth & $[2,3]$& $1$&$40$& Entanglement entropy estimation&MSE& 10& 100   &0.012\\
        Magnetic Phase Transition & $[2,11]$& 1   &$[5,32]$& Energy estimation&Arithmetic Mean& 
15& 100  &0.035\\ \bottomrule
    \end{tabular}
\label{table: performance}
\end{table*}

\section{II. Quantum flow matching}

QFM is designed to learn the stepwise evolution of a density matrix, illustrated by the time evolution of ring-shaped state ensembles (Fig.~\ref{fig:QFM}a). Starting from a density matrix $\rho_{0}$ approximated by an ensemble of $M$ sampled pure states $S_{0}=\{\ket{\psi_{0}^{1}},\ldots,\ket{\psi_{0}^{M}}\}$, QFM evolves each $\ket{\psi_{0}^{m}}$ independently over $\mathcal{T}$ steps with a fixed circuit, yielding a final ensemble $S_{\mathcal{T}}$ that approximates the target density matrix $\rho_{\mathcal{T}}$. At selected steps (Fig.~\ref{fig:QFM}b), ancilla qubits are inserted and measured, and the measurement outcomes condition the subsequent circuit evolution, thereby shaping the state distribution and realizing an explicit mapping between two ensembles. Such ensemble-to-ensemble transport, enabled by measurement-conditioned dynamics, is beyond conventional variational or ancilla-assisted circuits mainly used for single-state preparation or observable estimation. Quantum diffusion models such as QuDDPM~\cite{QuDDPM} correspond to the QFM framework when the initial ensemble is a Haar-random ensemble, all evolution steps are implemented through ancilla-assisted measurements, and the circuits are optimized using a fidelity-based loss function.

These measurement-based dynamics have attracted growing interest, including hybrid reconstruction of diffusion trajectories~\cite{liu2025measurement} and Petz-map recovery of ensemble averages~\cite{hu2025local}. QFM evolves arbitrary initial density matrices within a single reusable circuit architecture (Fig.~\ref{fig:QFM}c), enabling an ensemble-propagation framework for diverse applications (Fig.~\ref{fig:QFM}d) while avoiding sample-dependent circuit redesigns.

The circuit of QFM can be either analytically fixed or trainable, depending on target problems. For ensembles of Hamiltonians differing only in local terms, we construct circuits without training, whereas more general cases require a training protocol. We denote the circuit at step $\tau$ as $V(\vec{\theta}{\tau})$ with parameter vector $\vec{\theta}{\tau}$. Circuits with $n_a$ ancilla qubits are written as $V(\vec{\theta}{\tau})=U_{n_a}(\vec{\theta}{\tau})$, while those acting only on $n$ data qubits are $V(\vec{\theta}{\tau})=U_{n}(\vec{\theta}{\tau})$. For problems such as superdiffusion, $V(\vec{\theta}{\tau})$ can be chosen with analytically structures. By measuring ancilla qubits, QFM generates an ensemble of distinct evolutions with a single circuit.

For problems requiring a trainable circuit, such as estimating the free-energy change, QFM employs at each step an $L$-layer entanglement-varied hardware-efficient ansatz (EHA)~\cite{PhysRevApplied.21.034059} with trainable parameters $\vec{\theta}{\tau}$. Each layer applies gates $U^{(1)}=\prod_{j\in\{x,y,z\}}\exp\{-i\theta_{j}\sigma_j/2\}$ on each qubit, followed by two-qubit gates $U^{(2)}=\prod_{j\in\{x,y,z\}}\exp\{-i\theta_{j}\sigma_{j}\otimes\sigma_{j}/2\}$ sequentially on all nearest-neighbor pairs, where $\theta_{j}\in\vec{\theta}{\tau}$ and $\sigma_{j}$ are Pauli operators. When $V(\vec{\theta}{\tau})=U_{n}(\vec{\theta}{\tau})$, the EHA acts directly on the data qubits. When $V(\vec{\theta}{\tau})=U_{n_a}(\vec{\theta}{\tau})$, measuring the $n_a$-th ancilla qubit with outcome $r_{n_a}\in\{0,1\}$ is equivalent to inserting, between $U^{(1)}$ and $U^{(2)}$ on its neighboring data qubit in each layer, a single-qubit gate $U_{(r_{n_a})}=\exp\{-i\theta(-1)^{r_{n_a}}\sigma_{z}/2\}$. If $U^{(2)}$ is allowed to couple each ancilla qubit with data qubits rather than only nearest neighbors, $U_{(r_{n_a})}$ depends on all ancilla measurement outcomes, increasing both expressivity and circuit depth.

\begin{figure}[htp]
    \centering
    \includegraphics[width=0.8\linewidth]{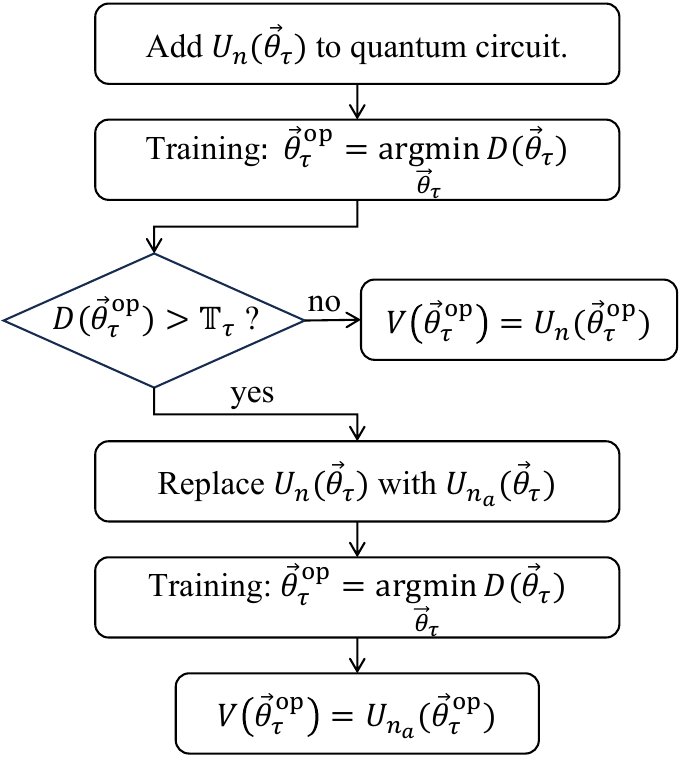}
 \caption{The training procedure of QFM in step $\tau$. QFM first adopts an $n$-qubit unitary operator for each layer; if the loss for $U_{n}$ does not converge below the threshold $\mathbb{T}_{\tau}$ during training, $n_a$ ancilla qubits are introduced and the training is repeated again.  }
    \label{fig:training}
\end{figure}

\begin{figure*}[htp]
    \centering
    \includegraphics[width=1\linewidth]{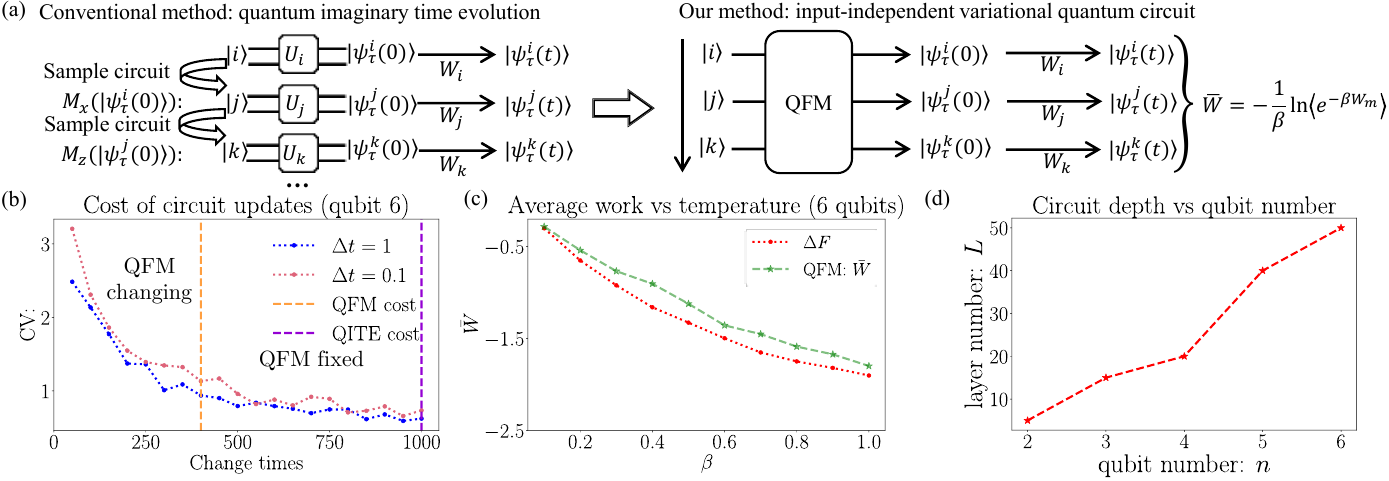}
 \caption{QFM efficiently estimates nonequilibrium free energy changes in a $6$-qubit TFIM. (a) Free energy is estimated via sampling pseudo-thermal states in METTS. Left:  The quantum imaginary time evolution (QITE)~\cite{QITE_1,QITE_2,Qu_Jar_exp_1} requires repeated circuit adjustments for each state. Right: QFM generates the ensemble with a fixed circuit after training. (b) For $\beta=1$, the coefficient of variation under time steps $\Delta t=1$ (blue) and $0.1$ (red) shows that QFM converges with $\sim400$ circuit updates (orange), compared with $\sim1000$ for QITE (purple)~\cite{Qu_Jar_exp_1}, with a larger gap for shorter $\Delta t$. (c) The averaged work agrees with the free-energy change versus inverse temperature $\beta$. (d) The QFM circuit depth scales nearly linearly with the qubit number.}
    \label{fig:sim_thermal}
\end{figure*}

\section{III. Generation and training}
Starting from an initial state $\ket{\psi_0^{m}} \in S_0$, QFM sequentially applies $V(\vec{\theta}_{1}^{\text{op}}) \cdots V(\vec{\theta}_{\tau-1}^{\text{op}})$ to generate $\ket{\widetilde{\psi}_{\tau-1}^{m}} \in \widetilde{S}_{\tau-1}$, where the tilde indicates results generated by QFM and the superscript of $\vec{\theta}_{\tau}^{\text{op}}$  denotes ``optimized''. QFM uses a hybrid approach to construct $V(\vec{\theta}_{\tau}^{op})$, which switches between two types of circuits. The first is a partially measured circuit, where $n_a$ ancilla qubits initialized by rotation operator $R_{y}(\theta)=\exp(-i\theta\sigma_{y}/2)$ to get $\ket{\tau}=R^{\otimes n_a}_{y}({\tau\pi}/{\mathcal{T}})\ket{0}^{\otimes n_a}$ are projectively measured along the $z$-axis after evolving $\ket{\widetilde{\psi}_{\tau}^{m}}=U_{n_a}(\vec{\theta}_{\tau}^{\text{op}})\ket{\tau}\ket{\widetilde{\psi}_{\tau-1}^{m}}\in \widetilde{S}_{\tau}$. The second is a unitary circuit, where we directly apply the circuit on $n$ data qubits.

For applications where the quantum circuit needs to be constructed through training, QFM begins by training from the first step $V(\vec{\theta}_{1}^{\text{op}}) $, and then proceeds step by step up to $ V(\vec{\theta}_{\mathcal{T}}^{\text{op}}) $. At each step, the circuit that have already been trained are fixed. Specifically, after completing the training up to step $\tau - 1$, we proceed to train the $V(\vec{\theta}_{\tau})$ at step $\tau$. We sequentially apply operators from $V(\vec{\theta}_{1}^{\text{op}})$ to $V(\widetilde{\theta}_{\tau-1}^{\text{op}})$, generating the state ensemble $\widetilde{S}_{\tau-1}=\{\ket{\widetilde{\psi}_{\tau-1}^{1}}, ...,\ket{\widetilde{\psi}_{\tau-1}^{M}}\}$. We then apply $U_{n}(\vec{\theta}_{\tau})$ to states sampled from $\widetilde{S}_{\tau-1}$ and minimize the loss function $D(\vec{\theta}_{\tau})$ to get the $\vec{\theta_{\tau}^{\text{op}}}$. If $D(\vec{\theta}_{\tau}^{\text{op}})$ is lower than the threshold $\mathbb{T}_{\tau}$, $U_{n}(\vec{\theta}_{l}^{\text{op}})$ will be the $\tau$-th step of QFM; otherwise, we switch $U_{n}(\vec{\theta}_{\tau}^{\text{op}})$ to $U_{n_a}(\vec{\theta}_{\tau})$ and repeat the same training process with input state $\widetilde{S}_{\tau}^{n_a}=\{\ket{\tau}\ket{\widetilde{\psi}_{\tau-1}^{1}},... ,\ket{\tau}\ket{\widetilde{\psi}_{\tau-1}^{M}}\}$ as shown in Fig.~\ref{fig:training}. To avoid barren plateaus, we initialize parameters in each layer with a neighborhood of radius $\mathcal{O}(1/L)$~\cite{grant2019initialization}. We determine the $\tau$-th step parameter $\vec{\theta}_{\tau}$ by minimizing the loss function, which is generically constructed to quantify the discrepancy of the application-dependent observable statistics between the evolved ensemble at step $\tau$ and   the target ensemble:

\begin{align}
    E_{m}&=O_{\tau}(\ket{\psi_{\tau}^{m}},V(\vec{\theta}_{\tau})\ket{\tau}\ket{\widetilde{\psi}_{\tau-1}^{m}}), \\
\vec{\theta}^{\text{op}}_{\tau}&=\min_{\vec{\theta}_{\tau}}D(\vec{\theta}_{\tau})=\min_{\vec{\theta}}f_{\tau}( E_{1},E_{2},...,E_{M}),    
\end{align}
where $O$ and $f$ are the operator and the function depending on the application. In general, the loss is chosen to compare a set of measurable ensemble observables whose expectation values sufficiently characterize the target distribution at each time step, so that matching these observables enforces the desired flow of the ensemble toward the target density matrix. For examples requiring the fidelity estimation, $O$ is the swap-test operator and $f$ is the average function. For the example of entanglement growth, $O$ estimates the entanglement entropy of $\ket{\widetilde{\psi}_{\tau}^{m}}$ and $f$ is the mean squared error (MSE). Training introduces an additional overhead in QFM through the optimization of circuit parameters. For applications requiring many ensemble samples, this cost can be distributed over repeated use of the trained circuit.

\begin{figure*}[htp]
    \centering
    \includegraphics[width=1\linewidth]{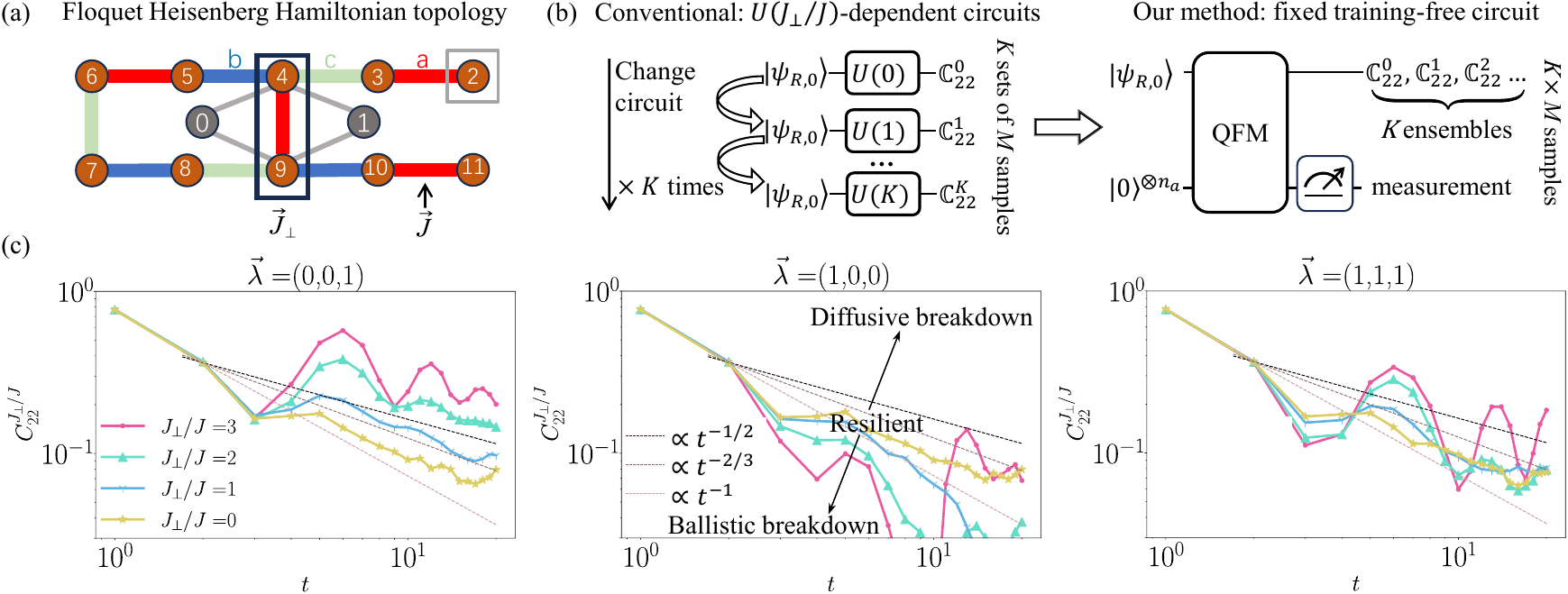}
 \caption{QFM facilitates the estimation of superdiffusive scaling. (a) The conventional method~\cite{kumaran2025quantum} requires adjusting the quantum circuits for a Heisenberg model under various 2D-interaction strengths (black box). (b) QFM evolves the state ensemble with a fixed circuit and adjusts the 2D-interaction strength by measurements on ancilla qubits (gray node in (a)). (c) The diffusive behaviors by QFM for 2D-interaction strengths $J_{\bot}/J\in\{0,1,2,3\}$ in three interaction types, matching the conventional method~\cite{kumaran2025quantum}. Increasing $J_{\bot}/J$ accentuates both diffusive ($\vec{\lambda} = (0,0,1)$) and ballistic ($\vec{\lambda} = (1,0,0)$) breakdown. In the resilient case ($\vec{\lambda} = (1,1,1)$), stronger 2D interactions enlarge the variance of the superdiffusive scaling. }
    \label{fig:sim_diff}
\end{figure*}

\section{IV. Applications}
We demonstrate the method on representative tasks. The benchmarks comparing QFM with QuDDPM are summarized in the main text, with further details provided in the Appendix. The main text focuses on free-energy estimation (variational circuit) and superdiffusive scaling (fixed circuit).

\subsection{A. Benchmark tasks }
To benchmark QFM against QuDDPM, we consider three representative examples. The ring-state evolution illustrates QFM's capability to map between two quantum state ensembles beyond the Haar-random initialization of QuDDPM. QFM also enables the generation of states with controlled entanglement entropy, the growth from separable to maximally entangled states, and the learning of the evolution from an ordered ferromagnetic phase to a disordered phase, capturing the phase transition. In each example, we evaluate the performance of QFM by the mean Kullback–Leibler (KL)-divergence~\cite{carrasquilla2017machine} between the generated and target distributions of an observable during the time evolution. The observable depends on the task: measurement results in the $Y$-direction for ring-state evolution, entanglement entropy for entanglement growth, and magnetization for the magnetic phase transition. We consider the KL-divergence below $0.05$ to indicate reliable distributional matching, which in practice also corresponds to a high state fidelity ($99.5\%$) between the generated and target ensembles. Alternative metrics such as the Hellinger distance~\cite{roga2016geometric} or trace distance provide qualitatively consistent results. A summary of the performance is provided in Table~\ref{tab:comparison}.

\subsection{B. QFM facilitates free energy estimation}
Having established the overall performance through benchmark comparisons, we now turn to two central tasks. We first consider QFM for facilitating free-energy estimation. The celebrated Jarzynski equality~\cite{Cl_Jar_1} enables us to calculate the free energy change between two equilibrium states, from the work in the nonequilibrium process. In quantum simulations, the quantum Jarzynski equality~\cite{Qu_Jar_1,Qu_Jar_2,Qu_Jar_3,Qu_Jar_4} is typically tested by computing the average work $\bar{W}$ from evolving METTS~\cite{Qu_Jar_exp_1,stoudenmire2010minimally,PhysRevLett.102.190601} under a time-dependent Hamiltonian $H(t)$. METTS prepares a thermal density matrix by sampling classical product states (CPS), e.g.,  $\ket{i}$, via alternating $M_x$ and $M_z$ measurements and projecting each sampled CPS independently toward the thermal state by QITE. In conventional QITE, the circuit must be reconfigured for each sample to achieve thermalization (Fig.~\ref{fig:sim_thermal}a, left), and real-time evolution must be carried out for every thermal sample in free-energy estimation, resulting in an experimental overhead that scales with the number of samples.

QFM generates the METTS ensemble with a single circuit (Fig.~\ref{fig:sim_thermal}a, right). Starting from $\ket{\psi_{0}^{m}}$, we apply $V(\vec{\theta}_{1}^{\mathrm{op}})\cdots V(\vec{\theta}_{\tau}^{\mathrm{op}})$ to obtain the target METTS $\ket{\widetilde{\psi}_{\tau}^{m}}$ at inverse temperature $\beta_{\tau}$, with the next input prepared by measurement. The parameters $\vec{\theta}_{\tau}^{\mathrm{op}}$ are obtained by minimizing the loss function:
\begin{equation}
    D(\vec{\theta}_{\tau})=(1/M)\sum_{m=1}^{M}\braket{\psi_{\tau}^{m}|V(\vec{\theta}_{\tau})|\widetilde{\psi}_{\tau-1}^{m}},
\end{equation}
where $S_\tau=\{\ket{\psi_{\tau}^m}=A^{-1}e^{-\beta_{\tau}H(0)/2}\ket{m}\}$ are training data prepared by QITE on CPS states $\ket{m}$, with $A=\sqrt{\langle\psi_{\tau}^m|e^{-\beta_{\tau}H(0)}|\psi_{\tau}^m\rangle}$, followed by real-time evolution to obtain $\ket{\psi_{\tau}^m(t)}$ (Appendix).

For a $6$-qubit TFIM $H(t)=\sum_i \sigma_z^i\sigma_z^{i+1}+g(t)\sum_i \sigma_x^i$ with $g(t)=1+t/20$ ($t=0\to10$), QFM generates METTS ensembles $\widetilde{S}\tau$ at inverse temperatures $\beta\tau=\tau/\mathcal{T}$ with $\mathcal{T}=10$ and $L=50$. In Fig.~\ref{fig:sim_thermal}b ($\beta=1$), once $\bar{W}$ converges, QFM reduces the total number of circuit adjustments by about $60\%$ compared with~\cite{stoudenmire2010minimally,PhysRevLett.102.190601}, with larger savings for faster dynamics. This benefit comes at the cost of a one-time training stage ($\sim400$ circuit updates), after which the same circuit can be reused across ensemble generation.

Performing the training classically further lowers this cost. QFM also reproduces the average work and the analytical free-energy change over temperatures (Fig.~\ref{fig:sim_thermal}c), and its circuit depth scales nearly linearly with the qubit number (Fig.~\ref{fig:sim_thermal}d).

\subsection{C. Superdiffusion}

We next consider superdiffusive scaling in fixed circuits, which provides a complementary setting to the variational scenario above. Quantum simulation was recently performed to  study superdiffusion~\cite{kumaran2025quantum}. QFM extends this study to explore how superdiffusion depends on the strength and type of interactions in the 2D Heisenberg model, as a Floquet extension of a 1D system. In the limit of vanishing 2D interaction strength $J_{\bot}$ (black box in Fig.~\ref{fig:sim_diff}a), the model reduces to an infinite-temperature 1D Heisenberg chain with interaction strength $J$. The first-order Trotterization with $\mathcal{T}$ steps is $e^{-iH_{2D}(t)} = (e^{-iH_a t/\mathcal{T}} e^{-iH_c t/\mathcal{T}} e^{-iH_b t/\mathcal{T}})^{\mathcal{T}}$, where $H_k = \sum_{\langle i,j \rangle \in k} h_{i,j}$ for $k \in \{a,b,c\}$ denotes the Hamiltonian with bonds of type $k$ and $h_{i,j}(\vec{J})=(J/4)\sum_{j=x,y,z}\lambda_j\sigma_j^i\sigma_j^j$ with $\vec{\lambda}=(\lambda_x,\lambda_y,\lambda_z)$. The 1D couplings are $\vec{J}=(1,1,1)\times J$, while 2D couplings are $\vec{J}_{\bot}=\vec{\lambda}\times J_{\bot}$. For the 10-qubit system (Fig.~\ref{fig:sim_diff}a), the 1D bonds of type $a$ are $\langle2,3\rangle$, $\langle5,6\rangle$, $\langle10,11\rangle$, of type $c$ are $\langle3,4\rangle,\langle6,7\rangle,\langle8,9\rangle$, of type $b$ are $\langle4,5\rangle,\langle7,8\rangle,\langle9,10\rangle$, and the 2D bond is $\langle4,9\rangle$ in  type $a$.

The conventional simulations on superdiffusion scaling require $K$ circuits, each of which evaluates the correlation function $C_{22}^{J_{\bot}/J}$ at distinct $J_{\bot}/J$ values  with $M$ random states $\ket{\psi_{R,0}}$. The correlation function can be expressed as a single-point observable~\cite{PhysRevLett.126.230501}: 
\begin{align}
    C_{pp}(t)=\frac{1}{2M}\sum \braket{\psi_{R,p}|\sigma_{z}^{p}(t)|\psi_{R,p}},
\end{align}
where $\ket{\psi_{R,p}}=\ket{0}_{p}\ket{\psi_{R}}$, the $\ket{\psi_{R}}$ is a Haar-random state, and $p$ is the location of the probe qubit. In the example of Fig.~\ref{fig:sim_diff}, the correlation function is evaluated using 100 states per interaction strength, with $p = 2$ and $M=100$ for each interaction strength.

In this example, we consider $J_{\bot}/J\in\{0,1,2,3\}$ and $\mathcal{T}=20$. The quantum circuit is $U(J_{\bot}/J)=(U_{1D}(a)U_{2D}(a)U_{1D}(c)U_{1D}(b))^{\mathcal{T}}$, where $ U_{1D}(a)=\prod_{\text{a type}}\exp\{-ih_{i,j}(\vec{J})\}$ and $ U_{2D}(a)=\exp\{-ih_{4,9}(\vec{J})\}$. It requires repeated circuit adjustments that increase the experimental overhead (Fig.~\ref{fig:sim_diff}b left). QFM realizes the same task with a single fixed circuit (Fig.~\ref{fig:sim_diff}b right), without a substantial circuit depth increase. It introduces two ancilla qubits (grey nodes in Fig.~\ref{fig:sim_diff}a) and adjusts the 2D-interaction strength through measurements, enabling uniform $J_{\bot}/J$ sampling without circuit redesigns. The QFM circuit over steps $1$ to $\tau$ is
\begin{align}
U_{n_a}(\tau)=
\Big[&U_{1D}(a)\times\exp\{-i[\sigma_{z}^{0}+0.5\sigma_{z}^{1}]\otimes h_{4,9}(\vec{J}_{\bot})\}\notag\\
&\times\exp\{-i\times 1.5h_{4,9}(\vec{J}_{\bot})\} \notag\\
&U_{1D}(c)U_{1D}(b)\Big]^{\tau}\text{H}_{\text{gate}}^{0}\text{H}_{\text{gate}}^{1},
\end{align}
where $\text{H}_{\text{gate}}^{0}$ and $\text{H}_{\text{gate}}^{1}$ are Hadamard gates that act on qubit $0$ and $1$ to balance the probability of each measurement result. After measurements, the circuit becomes $U_{n_a}(\tau)=(U_{1D}(a)U_{2D}(a)U_{1D}(c)U_{1D}(b))^{\tau}$, where
\begin{align}
    U_{2D}(a)=&\exp\{-i[(-1)^{r_0}+0.5(-1)^{r_1}]\otimes h_{4,9}    (\vec{J}_{\bot})\}\notag\\
&\times\exp\{-i\times 1.5h_{4,9}(\vec{J}_{\bot})\},
\end{align}
and  $r_{i}$ ($i\in \{0,1\}$) is independently and uniformly sampled from $\{0,1\}$, yielding circuits with $J_{\bot}/J=0,1,2,3$. 

We show how interaction strengths affect superdiffusion 
in Fig.~\ref{fig:sim_diff}c. For $\vec{\lambda}=(0,0,1)$, the system exhibits diffusive breakdown at $J_{\bot}/J\ge 2$. For $\vec{\lambda}=(1,0,0)$, ballistic transport breaks down at $J_{\bot}/J\ge 1$. The system remains resilient for $\vec{\lambda}=(1,1,1)$. Thus, QFM achieves consistent results by a single circuit, eliminating the multiple circuits required by the conventional method~\cite{kumaran2025quantum}.

\section{V. Generalization error}
We analyze the bound on the generalization error~\cite{Error_cl,caro2023out} of QFM for training-based tasks. To learn a unitary $U_{\tau}\in\mathcal{U}$ at step $\tau$, we have a training set $S_{\tau}=\{(\ket{\psi_{\tau-1}^{m}},U_{\tau}\ket{\psi_{\tau-1}^{m}})\}_{m=1}^{M}$. The training cost is:
\begin{align}
        D(\vec{\theta}^{\text{op}}_{\tau})&=\frac{1}{4M}\sum_{m=1}^{M}\|U_{\tau}\ket{\psi^{m}_{\tau-1}}\bra{\psi^{m}_{\tau-1}}U_{\tau}^{\dagger}\notag\\&\quad-   V(\vec{\theta}^{\text{op}}_{\tau})\ket{\widetilde{\psi}^{m}_{\tau-1}}\bra{\widetilde{\psi}^{m}_{\tau-1}}V(\vec{\theta}^{\text{op}}_{\tau})^{\dagger}\|_{1}^{2}.
\end{align}
The distance between $V(\vec{\theta}_{\tau}^{\text{op}})$ and target unitary $U_{\tau}$ gives the expected risk:
\begin{align}
        R_{P_\tau}(\vec{\theta}_{\tau}^{\text{op}})&=\frac{1}{4}\mathbb{E}_{\ket{\Psi_{\tau-1}}\sim P}\Big[\Big\|U_\tau\ket{\Psi_{\tau-1}}\bra{\Psi_{\tau-1}}U^{\dagger}_\tau\notag\\&\quad-V(\vec{\theta}_{\tau}^{\text{op}})\ket{\Psi_{\tau-1}}\bra{\Psi_{\tau-1}}V(\vec{\theta}_{\tau}^{\text{op}})^{\dagger}\Big\|_{1}^{2}\Big],
\end{align}    
where $P_\tau$ is the test set at step $\tau$. For a $\mathcal{N}$-gate unitary layer $U_{n}(\vec{\theta}_{\tau}^{\text{op}})$ in QFM, the expected risk $R_{P}^{n}(\vec{\theta}^{\text{op}}_{\tau})$ obeys:
\begin{align}
    R_{P}^{n}(\widetilde{\theta})\leq  2D_{\tau}(\vec{\theta}^{\text{op}}_{\tau})+\mathcal{O}\left(\sqrt{\frac{\mathcal{N}\log{(\mathcal{N})}}{M}}\right).
\end{align}
For partially measured layer $U_{n_a}(\vec{\theta}_{\tau}^{\text{op}})$, the expected risk $R_{P}^{n_a}(\vec{\theta}^{\text{op}}_{\tau})$ satisfies:
\begin{align}
\mathbb{E}_{\tau}&=\frac{1}{M}\sum_{m=1}^{M}\left| \braket{\widetilde{\psi}_{\tau}^{m}|U_{\tau}|\psi_{\tau-1}^{m}}\right|^{2},\\
    D_{\tau}^{n_a}(\vec{\theta}^{\text{op}}_{\tau})&=1-\mathbb{E}_{\tau},\\
    R_{P}^{n_a}(\vec{\theta}_{\tau}^{\text{op}})&\leq 2 D_{\tau}^{n_a}(\vec{\theta}^{\text{op}}_{\tau})+\mathcal{O}\left(\sqrt{\frac{\mathcal{N_{\text{eff}}}\log{(\mathcal{N_{\text{eff}}})}}{M}}\right),
\end{align}
where $\mathcal{N}_{\text{eff}}$ is the number of gates in the $U_{n_a}$ without ancilla qubits.

\section{VI. Conclusion}

We have presented a quantum generative model QFM for transforming between two density matrices and tracking the time evolution of state distributions. Compared with the classical neural-network approaches~\cite{RevModPhys.91.045002,PhysRevLett.128.090501,tang2024learning,schmitt2025simulating} and quantum-classical models~\cite{PhysRevLett.102.190601,Qu_Jar_exp_1,zapusek2025variational}, QFM employs a single quantum circuit and provides more efficient estimations of observables. We have demonstrated QFM's capabilities in learning topological state evolution, magnetic phase‐transition simulation, and entanglement‐entropy growth. Especially, QFM can bypass costly circuit adjustments to estimate free energy changes and superdiffusive scaling, offering a promising approach to track nonequilibrium quantum dynamics with less experimental overhead, where the circuit depth scales nearly linearly with the number of qubits (Appendix). While current hardware may constrain fault-tolerant implementations on quantum processors, the advances in device performance~\cite{king2025beyond,wu2024boss,PhysRevLett.134.090601} and quantum error correction~\cite{acharya2024quantum,bravyi2024high,cao2025generative,zhang2025correcting} are expected to mitigate these constraints. Future directions include experimental implementations of QFM and its application to facilitate the simulation of a broader class of nonequilibrium quantum dynamics and quantum thermodynamics. 

\begin{table*}[htbp]
    \centering
    \caption{Representative diffusion models and flow matching models.}
    \label{tab:diffusion_flow_models}
    \begin{tabular}{lll}
        \toprule
        & Model / Method & Property\\
        \midrule
        \multirow{4}{*}{Diffusion}& DDPM (Denoising Diffusion Probabilistic Models)~\cite{DDPM_1}& Foundational model; noise addition and denoising\\
        & Score-based Generative Models (SGM)~\cite{DDPM_2}& Uses SDEs + score matching\\
        & DDIM (Denoising Diffusion Implicit Models)~\cite{song2020denoising}& Accelerated sampling via implicit reverse process \\
        & Latent Diffusion (LDM)~\cite{rombach2021high}& Diffusion in latent space; basis of Stable Diffusion \\
        \midrule
        \multirow{3}{*}{Flow Matching}& Flow Matching for Generative Modeling~\cite{Flow_matching_GM}& Sampling via ODE; general FM framework \\
        & Consistency Models~\cite{song2023consistency}& Refined training objective; faster convergence \\
        & Rectified Flow++ / Variants~\cite{guo2025variational}& Systematic FM framework; relation to diffusion \\
        \bottomrule
    \end{tabular}
\end{table*}

\section{Acknowledgments}
We acknowledge Lei Wang and Xiaopeng Li for helpful discussions. We thank Yi-Zhuang You for bringing the recent works~\cite{liu2025measurement,hu2025local} to our attention. 
This work is supported by Projects 12322501, 12575035 of the National Natural Science Foundation of China, and 2026NSFSCZY0124 of the Natural Science Foundation of Sichuan Province. 
The computational work is supported by the Center for HPC, University of Electronic Science and Technology of China. The code of this study is available from the corresponding author upon reasonable request.

\section*{Data availability}

The data that support the findings of this article are openly available~\cite{code}.

\section{APPENDIX A: FLOW MATCHING AND GENERATIVE DIFFUSION MODELS}
We review two major paradigms in generative modeling—flow matching and diffusion models—and highlight their respective mechanisms and properties, which provide a basis for understanding the Quantum Flow Matching (QFM) and its distinction from quantum diffusion models. The flow matching~\cite{Flow_matching_GM} learns a continuous vector field \(v(x,t)\) that transports samples \(x_t\) from any source distribution \(p_0(x)\) to any target distribution \(p_1(x)\) by solving the ordinary differential equation (ODE):
\begin{equation}
    \frac{dx_t}{dt} = v(x_t, t), \quad x_0 \sim p_0(x),
\end{equation}
with the objective of minimizing the mismatch between the pushed-forward prior and the target distribution along the trajectory. 

Differently, generative diffusion models~\cite{DDPM_1,DDPM_2}, described by a stochastic differential equation (SDE), define a forward stochastic process that gradually adds noise to the data and learns a reverse denoising process \(\hat{p}(x_{t-1}|x_t)\) from Gaussian noise to the target, typically optimized via the variational objective:
\begin{equation}
\mathcal{L}_{\mathrm{diffusion}} = \mathbb{E}_{x_0, \epsilon, t} \left[ \|\epsilon - \epsilon_\theta(x_t, t)\|^2 \right],
\end{equation}
where $\epsilon$ is the noise in this step. The key difference lies in the generation mechanism: Diffusion generates samples stochastically, mainly starting from Gaussian noise, and typically requires many denoising steps to handle complex or multimodal distributions. The flow matching is deterministic and can map between arbitrary distributions, enabling direct interpolation, exact likelihood evaluation, and transformation between diverse data types.  The flow matching can efficiently transform samples between different image classes, interpolate between embeddings in text or speech models, or morph between distinct multi-modal datasets—tasks that diffusion models cannot perform directly.  We list in the  Table~\ref{tab:diffusion_flow_models} the representative generative diffusion  models and flow matching models. 

Similarly, in the quantum regime, QFM is designed to offer efficient interpolation between state ensembles of two approximate density matrices, rather than requiring to start from a Haar-random ensemble as in quantum diffusion models, e.g., the Quantum Denoising Diffusion Probabilistic Model~\cite{QuDDPM}. 

\begin{figure*}[!htp]
    \centering
    \includegraphics[width=1\textwidth]{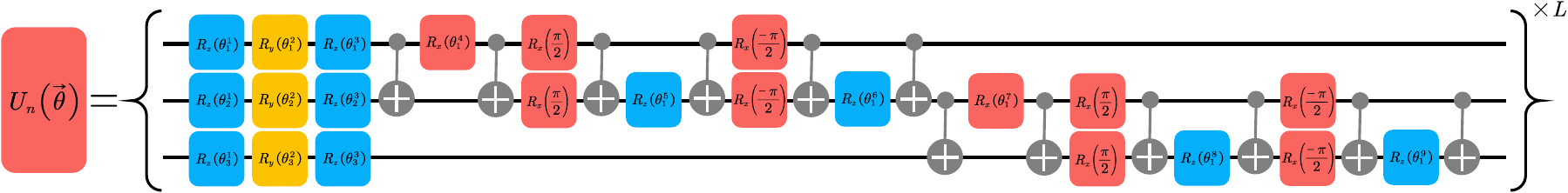}
    \caption{The unitary circuit $U_{n}(\vec{\theta})$ in QFM.  For steps without ancilla qubits in QFM, we apply the EHA circuit on data qubits. EHA  includes layers of $X$ and $Y$ single-qubit rotations followed by nearest-neighbor gates $XX$, $YY$, and $ZZ$, which provide tunable entanglement in the $x$, $y$, and $z$ directions.}
    \label{fig:sup_U_design}
\end{figure*}
\begin{figure*}[!htp]
    \centering
    \includegraphics[width=1\textwidth]{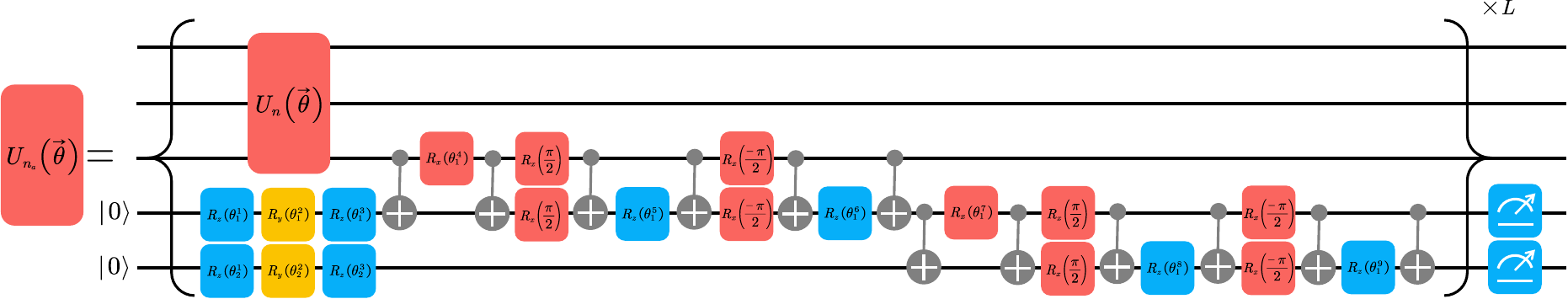}
    \caption{The circuit structure of steps with $n_a$ ancilla qubits in QFM.  For $U_{n_a}$, we add $n_a$ ancilla qubits, which are coupled to the data qubits through the same single- and two-qubit gates as in the EHA circuit, and measure the ancilla qubits with the Pauli-$Z$ operator at the end of each step.}
    \label{fig:sup_Una_design}
\end{figure*}

\section{APPENDIX B: QUANTUM DENOISING DIFFUSION PROBABILISTIC MODEL}

In this section, we give a brief introduction to QuDDPM. Inspired by DDPM, QuDDPM operates entirely within the quantum information processing paradigm and consists of two coupled quantum processes: a forward noising diffusion and a reverse denoising generation. (1) The forward diffusion process starts from a set of $n$-qubit quantum states $\mathcal{S}_0 = \{|\psi^{(0)}_i\rangle\}$ sampled from the target distribution $\mathcal{E}_0$. QuDDPM progressively injects noise by applying random local unitary operations (i.e., quantum scrambling circuits): $|\psi^{(k)}_i\rangle = \left( \prod_{\ell=1}^k U^{(i)}_\ell \right) |\psi^{(0)}_i\rangle, \quad k = 1, 2, \dots, T,$ where $U^{(i)}_\ell$ denotes a randomly chosen local unitary gate applied to the $i$-th sample at step $\ell$. After $T \sim \mathcal{O}(n / \log n)$ steps, the resulting ensemble $\mathcal{S}_T = \{|\psi^{(T)}_i\rangle\}$ becomes indistinguishable from a Haar-random (fully mixed) distribution. (2) The reverse denoising process starts from a noise ensemble $\tilde{\mathcal{S}}_T$ by sampling initial states $|\tilde{\psi}^{(T)}_i\rangle$ and iteratively apply a trainable parameterized quantum circuit (PQC) to reverse the diffusion: $|\tilde{\psi}^{(k-1)}_i\rangle \leftarrow \mathrm{Measure}\left[ \tilde{U}_k \left( |\tilde{\psi}^{(k)}_i\rangle \otimes |0\rangle^{\otimes n_A} \right) \right],$ where $\tilde{U}_k$ is a learnable unitary acting on the system qubits and $n_A$ ancillary qubits. A computational-basis measurement on the ancillas projects the output back onto a pure state, enabling a contractive mapping that gradually recovers structure from noise.

\subsection{1. Layerwise training strategy}
QuDDPM employs a hierarchical training scheme: at training stage $k$, the layers $\tilde{U}_{k+1}, \dots, \tilde{U}_T$ are frozen, and only $\tilde{U}_k$ is optimized to minimize the discrepancy between the generated ensemble $\tilde{\mathcal{S}}_k$ and the corresponding noisy ensemble $\mathcal{S}_k$ obtained from the forward process. This decomposes the overall circuit—whose depth would otherwise scale as $\Omega(n)$—into $T$ shallow sub-circuits of depth $\mathcal{O}(\log n)$, thereby mitigating barren plateaus and enabling stable optimization. The loss function of QuDDPM relies on distance measures between the generated and target ensembles. Two primary metrics are used. (1) Maximum Mean Discrepancy (MMD) with fidelity kernel $F(|\phi\rangle, |\psi\rangle) = |\langle \phi | \psi \rangle|^2$: $  D_{\mathrm{MMD}}(\mathcal{E}_1, \mathcal{E}_2) = \bar{F}(\mathcal{E}_1, \mathcal{E}_1) + \bar{F}(\mathcal{E}_2, \mathcal{E}_2) - 2\bar{F}(\mathcal{E}_1, \mathcal{E}_2),$   where $\bar{F}(\mathcal{E}_1, \mathcal{E}_2) = \mathbb{E}_{|\phi\rangle \sim \mathcal{E}_1,\, |\psi\rangle \sim \mathcal{E}_2} \big[ |\langle \phi | \psi \rangle|^2 \big]$. This expectation can be estimated experimentally via the Swap Test. (2) Wasserstein Distance: For data with nontrivial geometric or topological structure, we adopt the optimal transport-based Wasserstein distance of order $p$: $ W_p(\mathcal{E}_1, \mathcal{E}_2) = \left( \inf_{\pi \in \Pi(\mathcal{E}_1, \mathcal{E}_2)} \int |\langle \phi | \psi \rangle|^2 \, d\pi(|\phi\rangle, |\psi\rangle) \right)^{1/p},$ where $\Pi(\mathcal{E}_1, \mathcal{E}_2)$ denotes the set of all joint distributions with marginals $\mathcal{E}_1$ and $\mathcal{E}_2$.

\begin{figure*}
    \centering
    \includegraphics[width=1\textwidth]{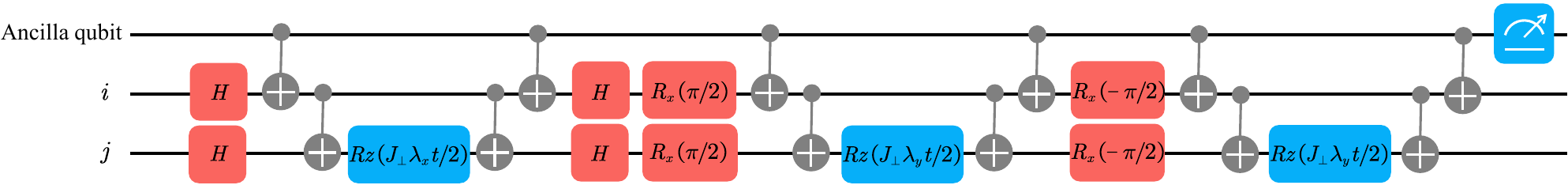}
    \caption{The Trotter circuit for the time evolution of 2D term in $U_{n_a}$ for time $t$.}
    \label{fig:an_hij}
\end{figure*}

\section{APPENDIX C: CIRCUIT STRUCTURE OF QUANTUM FLOW MATCHING}
In this section, we describe circuit designs of QFM, including the entanglement-varied hardware-efficient ansatz (EHA)~\cite{PhysRevApplied.21.034059} for training tasks and specialized training-free circuits for simulating the 2D Heisenberg model under superdiffusive scaling. We explain how ancilla measurements influence the circuit behavior and illustrate with the superdiffusive scaling how measurements on ancilla qubits can effectively modify the 2D interaction strength of the simulated system. This provides a clear understanding of both training-based and training-free QFM circuit design principles. 

\subsection{1. Circuit for variational quantum circuit}
In this subsection, we describe the $L$-layer EHA circuit in QFM for training tasks, highlighting its structure with and without ancilla qubits. The EHA circuit employs variational entanglers composed of parameterized  $XX$, $YY$, and $ZZ$ gates. This design allows the circuit to rapidly adjust the amount of entanglement to match the target ground state, thereby overcoming limitations such as barren plateaus and poor trainability. Compared with widely used ansatzes, including HEA (hardware-efficient ansatz )~\cite{kandala2017hardware}, HVA (Hamiltonian variational ansatz)~\cite{PRXQuantum.1.020319}, HSA (hardware symmetry-preserving ansatz)~\cite{Lyu2023symmetryenhanced}, and chemically inspired ansatzes such as UCCSD (unitary coupled cluster with all single and double excitations)~\cite{Romero_2019} and GRSD (Givens rotations with all single and double excitations)~\cite{Arrazola2022universalquantum}, EHA consistently achieves higher accuracy and greater robustness across both quantum many-body systems and molecular simulations. Importantly, EHA is problem-agnostic, making it broadly applicable without requiring specific prior knowledge of the Hamiltonian or tailored initial states.

Each layer of the EHA circuit includes arbitrary single-qubit rotations $U^{(1)}=\prod_{j\in\{x,y,z\}}\exp\{-i\theta_{j}\sigma_j/2\}$ and nearest-neighbor two-qubit gates $U^{(2)}=\prod_{j\in\{x,y,z\}}\exp\{-i\theta_{j}\sigma_{j}\otimes\sigma_{j}/2\}$, which we refer to as the $XX$, $YY$, and $ZZ$ gates depending on the choice of $j$. For intuitive understanding, we present in  Fig.~\ref{fig:sup_U_design} the  $L$-layer circuit structure of $U_{n}(\vec{\theta})$  for a three-qubit example, where each layer consists of arbitrary single-qubit gates and nearest-neighbor entangling gates along the $X$, $Y$, and $Z$ directions. All $\theta$ parameters in the figure are trainable.

In  Fig.~\ref{fig:sup_Una_design}, we illustrate how $2$ ancilla qubits are coupled to $3$ data qubits. After applying $U_n(\vec{\theta})$ on the data qubits, we first apply the single-qubit gates from the EHA circuit to each ancilla qubit. Then, on the ancilla qubit adjacent to the data qubits, sequentially apply the $XX$, $YY$, and $ZZ$ gates. The remaining ancilla qubits are similarly coupled sequentially to their nearest-neighbor ancilla qubits to construct the $U_{n_a}$.b At the end of $U_{n_a}$, each ancilla qubit is measured in the Pauli-$Z$ basis. 

\begin{figure}[!htp]
    \centering
    \includegraphics[width=1\linewidth]{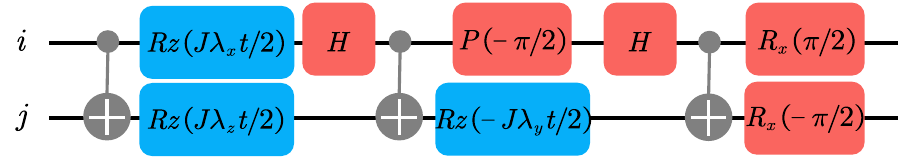}
    \caption{The Trotter circuit for time evolution of $h_{i,j}$ over time $t$.}
    \label{fig:hij}
\end{figure}

\begin{figure*}[!htpb]
    \centering
    \includegraphics[width=1.0\textwidth]{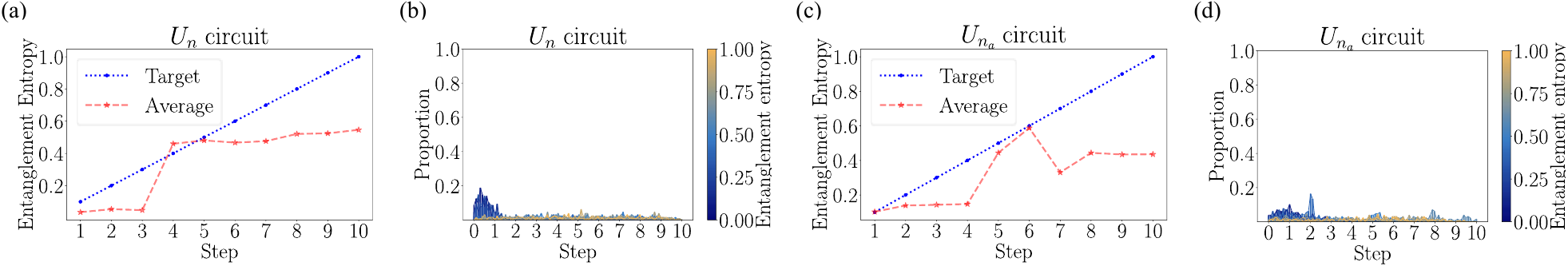}
    \caption{The learning performance in entanglement growth for each step of QFM with only $U_n$ and for each step of QFM with only $U_{n_a}$. Both circuits fail to learn the expected evolution process: the average entanglement entropy (red dashed line) does not match the target entanglement entropy (blue dashed line), and the entanglement entropy distribution of generated states at each step does not concentrate around the target entanglement entropy. }
    \label{fig:sup_comp_Un_Una}
\end{figure*}

\subsection{2. Circuit for 2D Heisenberg model }
In this subsection, we focus on the structure of QFM circuits in training-free tasks, describing how the 1D and 2D interaction components, along with ancilla qubits, are arranged to simulate the Heisenberg model and control interaction strengths without training. The 1D isotropic Heisenberg chain, an integrable model, displays superdiffusive spin transport governed by Kardar–Parisi–Zhang (KPZ) universality. In real materials~\cite{PhysRevLett.133.256301} and simulators, however, integrability-breaking interactions suppress this behavior, leading to a crossover to conventional diffusion or ballistic transport. Resolving this breakdown is key to connecting idealized models with quantum materials. 

We employ QFM to probe how two-dimensional (2D) interactions, native to scalable hardware architectures like IBM’s heavy-hex lattice, drive the suppression of superdiffusion. By tuning the strength and symmetry of these 2D couplings, we elucidate their role in destabilizing KPZ scaling. QFM employs a training-free circuit to simulate the 2D Heisenberg model with varying interaction strengths. By coupling ancilla qubits, the 2D interactions are adjusted based on their measurement outcomes. The circuit consists of a 1D interaction part $U_{1D}$ and a 2D interaction part $U_{2D}$.

We employ measurements on the ancilla qubits in QFM to modify the 2D interaction strength of the simulated Heisenberg system on data qubits. For clarity, we decompose the evolution operator into interaction operators along different bonds and dimensions. Since these interactions are local and mutually commuting, the first-order Trotter circuit for simulating the 2D Heisenberg model is:
\begin{equation}
     U(J_{\bot}/J)=(U_{1D}(a)U_{2D}(a)U_{1D}(c)U_{1D}(b))^{\mathcal{T}}.
\end{equation}
In $U_{1D}$, the Hamiltonian consists of a series of two-local terms  $h_{i,j}$, each acting on qubits $i$  and $j$. Since all these two-local terms mutually commute, the evolution under $U_{1D}$ can be factorized such that each term is implemented independently. Specifically, each two-local term  $h_{i,j}$ can be represented by a corresponding evolution operator $e^{-ih_{i,j}t}$, allowing the full unitary $U_{1D}$ to be expressed as the product of these operators. This decomposition facilitates efficient circuit implementation and ensures exact simulation of the 1D interactions.

\begin{equation}
    e^{-ih_{i,j}(\vec{J})t}=\exp\Big\{-iJ(\lambda_{x}\sigma_{x}^{i}\sigma_{x}^{j}+\lambda_{y}\sigma_{y}^{i}\sigma_{y}^{j}+\lambda_{z}\sigma_{z}^{i}\sigma_{z}^{j})t/4\Big\}.
\end{equation}

We implement the operator using the first-order Trotter–Suzuki decomposition, as shown in  Fig.~\ref{fig:hij}. For the 2D interaction term $U_{2D}$, QFM adjusts the interaction strength based on measurements of ancilla qubits. In the main-text example, the coupling strength between each ancilla qubit and the 2D system is parameterized by $\theta_{i}$, with an additional drift term $C$ applied to the 2D interaction: 
\begin{align}
U_{n_a}(\tau)=
\Big[&U_{1D}(a)\times\exp\{-i\frac{J_{\bot}}{4}[\theta_{0}(-1)^{r_0}+\theta_{1}(-1)^{r_1}+\notag\\
&\theta_{2}(-1)^{r_2}+\theta_{3}(-1)^{r_3}]\otimes h_{4,9}\}\notag \\
&\times\exp\{-i\frac{C}{4}J_{\bot}\times h_{4,9}\}U_{1D}(c)U_{1D}(b)\Big]^{\tau}.
\end{align}

To implement different 2D interaction strengths $J_{\perp}/J = 0, 1, 2, 3$ conditioned on outcomes of ancilla qubit measurements, we adjust the circuit parameters accordingly. Specifically, for the 1D interaction strength $J = 1$, the single-qubit rotation angles $\theta_i$ and the coupling constant $C$ are set to values that realize the desired interaction strength in the effective Hamiltonian. These parameter settings ensure that each measurement outcome on the ancilla qubits correctly tunes the two-dimensional coupling between data qubits. For the example in the main text, these parameters are set to:
\begin{equation}
\left\{
\begin{split}
\theta_{0} &= 1,\\
\theta_{1} &= 0.5,\\
\theta_{2} &= \theta_{3} = 0,\\
C &= 1.5.
\end{split}
\right.
\end{equation}

Furthermore, to avoid additional sampling overhead, it is necessary to generate a uniform distribution over $J_{\bot}/J$. This can be achieved by preparing each ancilla qubit in an equal superposition of its computational basis states, which is analogous to applying a Hadamard gate  $\text{H}_{\text{gate}}$ to each ancilla qubit. In this way, all possible interaction strengths are sampled with equal probability without additional sampling or post-selection: 
\begin{align}
U_{n_a}(\tau)=
\Big[&U_{1D}(a)\times\exp\{-i[\sigma_{z}^{0}+0.5\sigma_{z}^{1}]\otimes h_{4,9}(\vec{J}_{\bot})\}\notag\\
&\times\exp\{-i\times 1.5h_{4,9}(\vec{J}_{\bot})\} \notag\\
&U_{1D}(c)U_{1D}(b)\Big]^{\tau}\text{H}_{\text{gate}}^{0}\text{H}_{\text{gate}}^{1},
\end{align}
where the first-order Trotter–Suzuki decomposition of the 2D interaction terms in $U_{n_a}$ is illustrated in  Fig.~\ref{fig:an_hij}, and interactions along different bonds and directions are implemented sequentially. After measurements, the circuit will be:
\begin{align}
    U_{n_a}(\tau)=&(U_{1D}(a)U_{2D}(a)U_{1D}(c)U_{1D}(b))^{\tau},\\
    U_{2D}(a)=&\exp\{-i[(-1)^{m_0}+0.5(-1)^{m_1}]\otimes h_{4,9}    (\vec{J}_{\bot})\}\notag\\
&\times\exp\{-i\times 1.5h_{4,9}(\vec{J}_{\bot})\},
\end{align}
where each $r_{i}$ with $i\in \{0,1\}$ is sampled independently with equal probability from $\{0,1\}$. For the 2D interaction strength defined as $J_{\bot}=\theta_{a_0}(-1)^{r_0}+\theta_{a_1}(-1)^{r_1}+\theta_{a_2}(-1)^{r_2}+\theta_{r_3}(-1)^{r_3}+C$, the resulting distribution of $J_{\bot}$ is determined by the independent binary sampling of each $r_i$. This procedure ensures that all possible combinations of signs occur with equal probability, producing a discrete set of interaction strengths that can be used to implement a uniform or tailored ensemble of 2D couplings conditioned on the ancilla measurements:
\begin{align}
    &p(r_{0}=0,r_{1}=0)=0.25: J_{\bot}=3,\\ 
    &p(r_{0}=0,r_{1}=1)=0.25: J_{\bot}=2,\\ 
    &p(r_{0}=1,r_{1}=0)=0.25: J_{\bot}=1,\\ 
    &p(r_{0}=1,r_{1}=1)=0.25: J_{\bot}=0.
\end{align}

\begin{figure*}[!htp]
    \centering
    \includegraphics[width=0.8\textwidth]{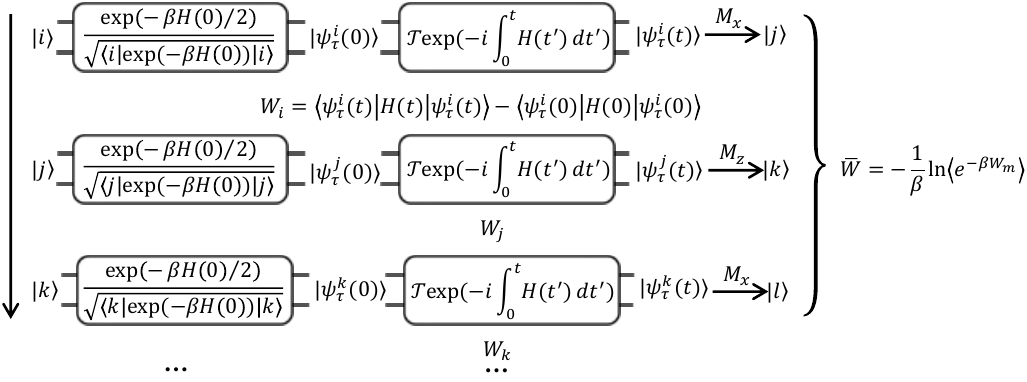}
    \caption{The procedure of the METTS algorithm at $\beta_{\tau}$. The procedure starts from classical product states (CPS) $|i\rangle$. Each CPS is evolved in imaginary time to obtain the thermal state $|\psi_\tau^{i}(0)\rangle$. Alternating projective measurements in the $z$ and $x$ bases generate the next CPS $|j\rangle$. The state $|\psi_\tau^{i}(0)\rangle$ is then evolved under the real-time dynamics to $|\psi_\tau^{i}(t)\rangle$, from which a work sample $W_i$ is obtained. The free-energy difference is finally estimated by averaging over the work samples.}
    \label{fig:METTS}
\end{figure*}

\subsection{3. Performance of non-adaptive circuits in entanglement growth}
In this subsection, we describe how the adaptive selection between $U_{n}$ and $U_{n_a}$ in QFM prevents disruptions to accurate and stable mapping of state distributions across steps. At step $\tau$, we adaptively select $U_{n}$ and $U_{n_a}$ to ensure convergency. $U_{n}$ handles shifts in the state distribution from step $\tau-1$, but cannot account for distortions, while $U_{n_a}$ can approximate distorted distributions. However, using $U_{n_a}$ on purely shifted distributions introduces fluctuations and hinders convergency. By appropriately choosing between $U_{n}$ and $U_{n_a}$, we accurately map the state distribution from step $\tau-1$ to $\tau$.

We numerically compare circuits using only $U_{n}$ with those using only $U_{n_a}$, e.g., QuDDPM, in generating entangled states. A circuit composed solely of $U_{n}$ cannot distort the distribution of quantum states; therefore, the average entanglement entropy of the generated states fails to match the growth of the target entanglement entropy ( Fig.~\ref{fig:sup_comp_Un_Una}a ). This is reflected in the distributions at each step: in the early steps, the entanglement entropy of the generated states exhibits a peak concentrated near the target entanglement entropy, but in later steps the distribution gradually becomes uniform (  Fig.~\ref{fig:sup_comp_Un_Una}b ). We speculate that this is because, at the beginning, the growth of entanglement entropy is relatively small, so the target state distribution is not far from the initial state distribution, allowing some degree of fitting. However, for highly entangled states, the shape of the target distribution differs significantly from the initial distribution, leading to increasingly poor learning performance in subsequent steps. In contrast, a circuit composed solely of $U_{n_a}$ tends to alter the shape of the state distribution at every step. This increases the training difficulty in steps where the distribution shape should ideally remain unchanged (  Fig.~\ref{fig:sup_comp_Un_Una}c ). Consequently, the generated states exhibit a large variance in entanglement entropy from the very beginning (  Fig.~\ref{fig:sup_comp_Un_Una}d ). This illustrates the limitation of using a single type of circuit and highlights the need for carefully choosing the circuit structure at each step. 

\begin{figure*}[htpb]
    \centering
    \includegraphics[width=1.0\textwidth]{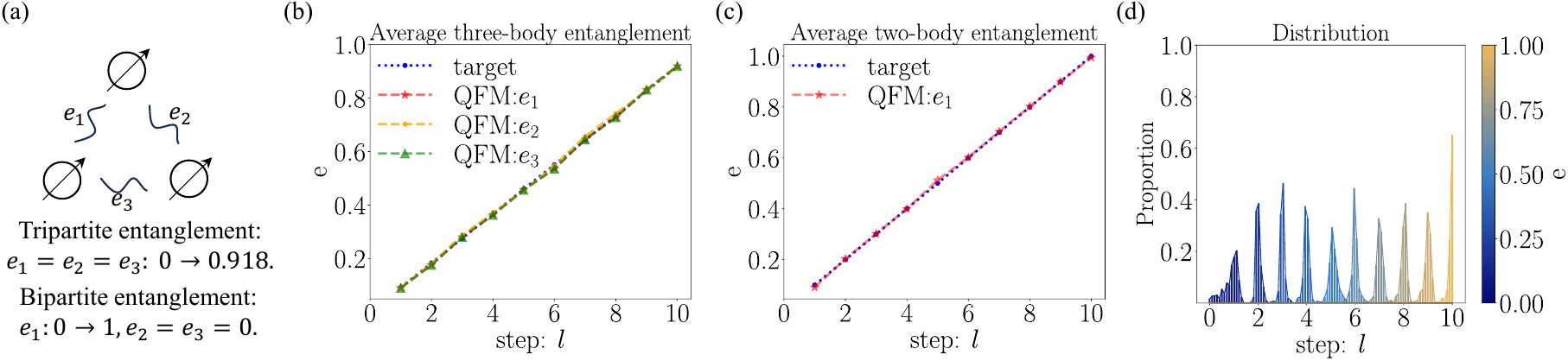}
    \caption{Generation of entangled state ensembles in the tripartite entanglement system and the bipartite entanglement system. (a) Experiments on tripartite and bipartite entangled systems. In the tripartite case, the entanglement entropy between any two qubits increases simultaneously at each step, reaching the tripartite upper bound of $e_1=e_2=e_3=0.918$. In the bipartite case, the entanglement between two selected qubits rises to the maximal value of $e=1$.  (b) Red, orange, and green dots represent the average entanglement entropy between each pair of qubits in the generated $3$-qubit state ensemble at various steps, while blue dots indicate the target entanglement entropy at each step. (c) Red dots are the average entanglement entropy of the generated $2$-qubit state ensemble at various steps, and blue dots are the target entanglement entropy in each step. (d) The entanglement entropy distribution of the $2$-qubit state ensemble generated at each step is denoted by different colors.} 
    \label{fig:sim_entangled}
\end{figure*}

\section{APPENDIX D: MEASUREMENT-BASED DYNAMICS}

In this section, we discuss and compare QFM with the Measurement-Based Quantum Diffusion Models (MBQDM)~\cite{liu2025measurement} under measurement-based dynamics~\cite{jacobs2006straightforward}. The measurement-based dynamics can be described by an SDE. For a pure state $\ket{\psi_t}$ at time $t$, continuously weakly measured by operator $X$ with measurement strength $k$, the stochastic evolution under weak measurements is governed by the stochastic Schrödinger equation:
\begin{align}
\mathrm{d}|\psi_t\rangle
&= \large[ -k\,(X_t-\langle X_t\rangle)^{2}\,\mathrm{d}t \notag\\
&+ \sqrt{2k}\,(X_t-\langle X_t\rangle)\,\mathrm{d}W_t \large] |\psi_t\rangle ,
\end{align}
where $\braket{X_t}=\braket{\psi_t|X_t|\psi_t}$ and $\text{d}W_t$ describes a standard Wiener process. In QFM, ancilla qubits are evolved and measured in the $\sigma_z$ basis to implement weak measurements in different bases, enabling states sampled from the initial ensemble to be mapped onto that of target states through an appropriate choice of measurement operators, realized via suitable evolution of the ancilla qubits. Meanwhile, MBQDM uniformly spreads states sampled from the initial ensemble across the Hilbert space by choosing $X_t$ from an appropriate distribution, forming the basis of quantum diffusion models. To further illustrate the difference between QFM and MBQDM in state recovery, we represent the initial state as the projection operator $\rho_t=\ket{\psi_t}\bra{\psi_t}$, and the stochastic partial differential equation takes the form: 
\begin{align}
\mathrm{d}\rho_t&= -k[X_t,[X_t,\rho_t]]\,\mathrm{d}t\notag\\
&+ \sqrt{2k}\,\big(X_t\rho_t+\rho_t X_t-2\langle X_t\rangle\rho_t\big)\,\mathrm{d}W_t.
\end{align}

MBQDM expands $\rho_{t}=\bm{z}_{t}\bm{P}/2^{n}$ and $X_{t}=\bm{x_{t}}\cdot\bm{P}$ in the Pauli-operator basis $\bm{P}=\{P_{i}\}_{i=1}^{4^{n}}$, with $z_{t,i}=\text{Tr}(\rho_{t}P_{i})=\braket{\psi_{t}|P_{i}|\psi_{t}}$ and $x_{t,i}=\text{Tr}(X_{t}P_{i})/2^{n}$. It adopts different strategies for state trajectory-level and ensemble-average recovery. For trajectory-level recovery, MBQDM adopts a classical–quantum hybrid scheme, where a classical decoder is trained to infer the Pauli-basis coefficient vector $\{\bm{z}_t\}$ from measurement operators and outcomes, and generate the parameter  $\theta$ of the denoising Hamiltonian $H_{\theta}(\bm{z}_t,t)$ for each state. For ensemble-average recovery, MBQDM applies the Petz recovery map to provide a principled way to invert quantum channels. Here, QFM generates an ensemble of distinct Hamiltonian evolutions from a single circuit by measuring the ancilla qubits. The evolution corresponding to each initial state is ensured, either through training or analytical methods, to evolve each state to its corresponding target. As a result, QFM achieves ensemble-average recovery within a single quantum circuit. 

\begin{figure*}[!htpb]
    \centering
    \includegraphics[width=1\textwidth]{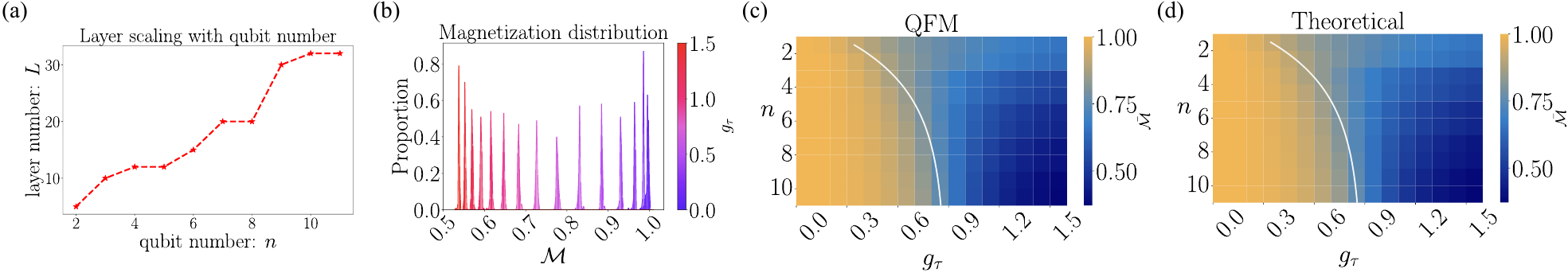}
    \caption{Generation of magnetized state ensemble. QFM is trained to generate state ensembles of different magnetization $\mathcal{M}$ under various magnetic field strengths $g$. (a) The number of QFM layers scales with the number of qubits. (b) Magnetization distribution of the ensemble generated at each step in a $4$-qubit system, denoted by color. (c) The transition line (white) and the average magnetization $\bar{\mathcal{M}}$ for various system sizes $n$ and $g$, which closely matches the theoretical result in (d).}
    \label{fig:sim_TFIM}
\end{figure*}

\section{APPENDIX E: ESTIMATION ON FREE ENERGY CHANGE}

In this section, we demonstrate how QFM can be used to estimate free-energy differences and verify the quantum Jarzynski equality. The Jarzynski equality~\cite{Cl_Jar_1} provides a remarkable connection between nonequilibrium processes and equilibrium free energy differences. It states that the exponential average of the work $W$ performed on a system during a finite-time, nonequilibrium process satisfies:
\begin{equation}
\langle e^{-\beta W} \rangle = e^{-\beta \Delta F},
\end{equation}
where \(\beta\) is the inverse temperature, \(\Delta F\) is the free energy difference between the initial and final equilibrium states, and the average is taken over many realizations of the process. This equality implies that equilibrium free energy differences can be obtained from measurements of work in arbitrarily far-from-equilibrium transformations, making it a powerful tool in both classical and quantum thermodynamics. It also underpins fluctuation theorems and has been widely applied in experimental and computational studies of small systems.

In the quantum computing~\cite{Qu_Jar_1,Qu_Jar_2,Qu_Jar_3}, minimally entangled typical thermal states (METTS)~\cite{stoudenmire2010minimally,PhysRevLett.102.190601} provide a conceptually simple approach to preparing finite-temperature quantum states by representing the thermal density matrix $\rho_{\beta}$ as an ensemble of pure states. In practice, each METTS sample is generated by starting from a classical product state (CPS)  $\ket{i}$, which is obtained through a sequence of projective measurements (typically alternating between the $M_z$ and $M_x$ bases).

To project such a CPS toward a thermal state, quantum imaginary-time evolution (QITE) is applied to approximate the non-unitary operator $\exp{-\beta H/2}$ and to obtain a normalized thermal-like state:
\begin{equation}
|\psi^{i}\rangle = \frac{e^{-\beta H/2}|i\rangle}{\| e^{-\beta H/2}|i\rangle \|}.
\end{equation}
Importantly, this QITE must be carried out independently for each sampled CPS, since different measurement outcomes lead to distinct initial states. As a result, the QITE circuit needs to be repeatedly adapted or re-optimized for every METTS sample in order to achieve proper thermalization ( Fig. \ref{fig:METTS}). Consequently, the total experimental overhead scales with the number of samples required to represent the thermal ensemble, making METTS costly to implement on near-term quantum devices where frequent circuit recompilation or parameter updates are challenging.

To estimate free energy differences using the Jarzynski equality, each METTS \(|\psi^m\rangle\) is evolved under a finite-time, nonequilibrium protocol corresponding to a Hamiltonian change, and the work \(W_m\) performed is recorded. The exponential average over the ensemble,
\begin{equation}
e^{-\beta\bar{W}}=\langle e^{-\beta W_m} \rangle = \frac{1}{M} \sum_{m=1}^{M} e^{-\beta W_m},
\end{equation}
then provides an estimate of the free energy difference $\Delta F$. This approach efficiently combines thermal-state sampling with quantum simulation of nonequilibrium dynamics, enabling accurate free energy estimation on quantum devices without requiring full thermal density matrix preparation. The thermal average work $\bar{W}$ is estimated by evolving sufficiently METTS. The detailed estimation procedure by QFM is as follows:

\begin{algorithm}[H]
\caption{Estimation of Thermal Average Work $\bar{W}$ via QFM}
\SetAlgoLined
\KwIn{Inverse temperature $\beta_{\tau}$, Hamiltonian $H(t)$}
\KwOut{Work samples $\{W_m\}$ and estimated $\Delta F$}
\For{each sample index $m = 1$ to $M$}{
  (1) QFM evolves a CPS state $\ket{i}$ into METTS $\ket{\widetilde{\psi}_{\tau}^m}=\ket{\widetilde{\psi}_{\tau}^m(0)}$\;
  (2) Compute initial energy: $E_i = \braket{\widetilde{\psi}_{\tau}^m(0) | H(0) | \widetilde{\psi}_{\tau}^m(0)}$\;
  (3) Apply real time-evolution to get: $\ket{\widetilde{\psi}_{\tau}^m(t)}$\;
  (4) Compute final energy: $E_{f}=\braket{\widetilde{\psi}_{\tau}^{m}(t)|H(t)|\widetilde{\psi}_{\tau}^{m}(t)}$\;
  (5) Calculate work sample: $W_m = E_f - E_i$\;
  (6) Measure $\ket{\widetilde{\psi}_{\tau}^m}$ to obtain next CPS state;
}
Compute thermal average: $e^{-\beta\Delta F} \approx e^{-\beta\bar{W}}=\braket{e^{-\beta W_{m}}}$.
\end{algorithm}
For the example in the main text, we set $M=1000$ for each $\beta_{\tau}$ in a $6$-qubit system.

\section{APPENDIX F: MORE BENCHMARKS FOR THE QUANTUM FLOW MATCHING}

In this section, we apply QFM to three quantum tasks—topological state evolution, entanglement growth, and the magnetic phase transition in the transverse-field Ising model (TFIM)—to demonstrate its versatility. Unlike QuDDPM, which generates specific target ensembles from a Haar-random ensemble, QFM can interpolate between different ensembles and capture entanglement growth, a task unattainable with the $U_{n_a}$ circuit alone.

\subsection{1. Ring state evolution}
\begin{figure}[!htp]
    \centering
    \includegraphics[width=1\linewidth]{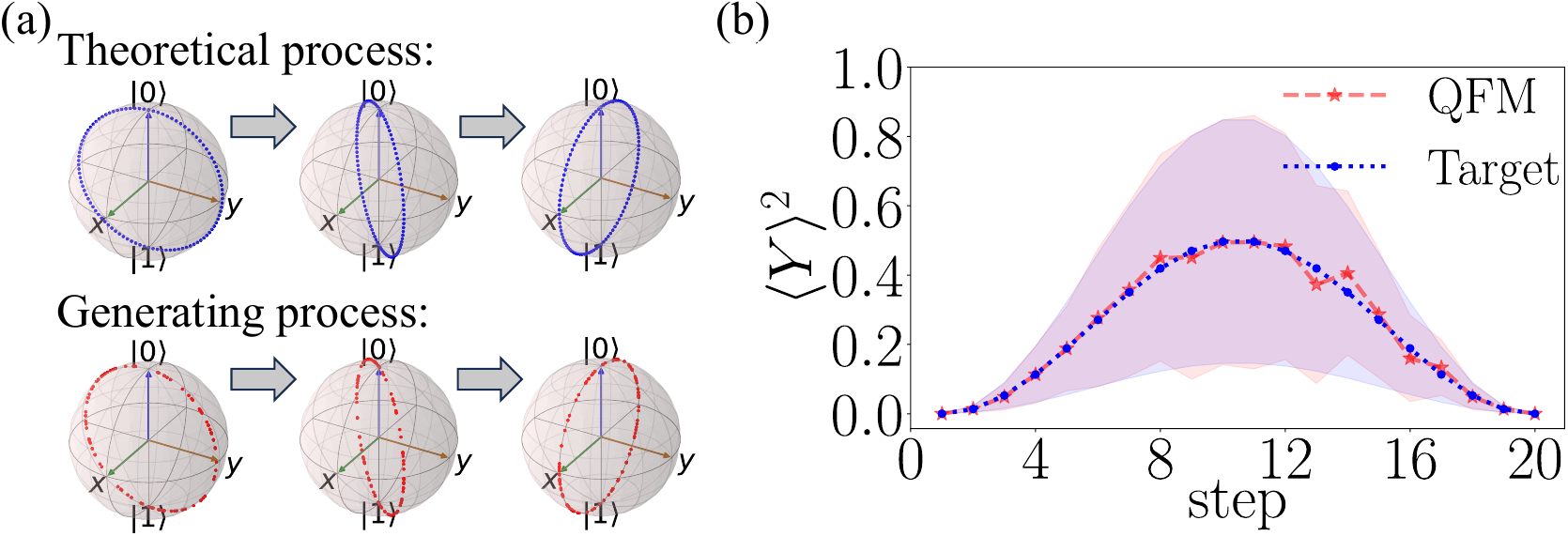}
    \caption{Generation of ring state ensemble. (a) Blue dots are sampled states used to train the model, and red dots are generated states. (b) The generated states (red) have matched deviation $\braket{Y^{2}}$ with the theoretical values (blue).} 
    \label{fig:sim_ring}
\end{figure}

\begin{figure*}[htp]
    \centering
    \includegraphics[width=1\textwidth]{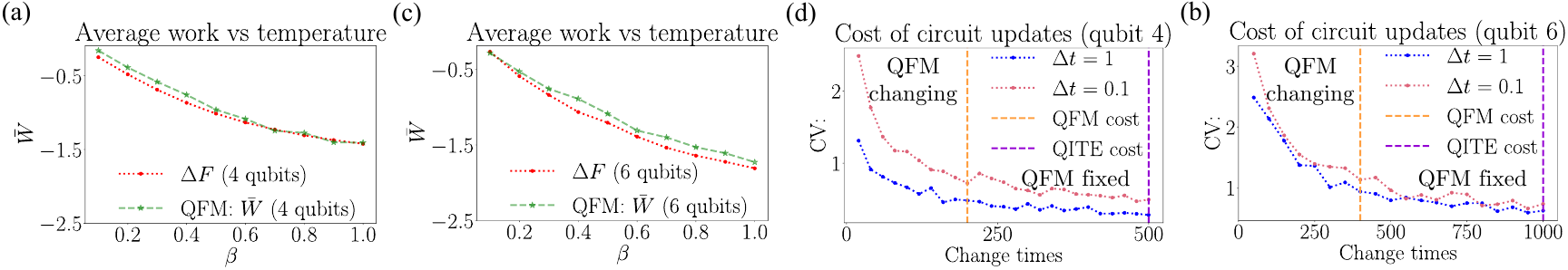}
    \caption{Free-energy estimation by QFM for four- and six-qubit systems. (a,c) For the four- and six-qubit cases, the averaged work agrees with the free-energy change as a function of the inverse temperature $\beta$.
(b,d) The coefficient of variation (CV) at $\beta=1$ for time steps $\Delta t=1$ (blue) and $\Delta t=0.1$ (red) shows that QFM converges with about 200 (four qubits) and 400 (six qubits) circuit updates (orange), whereas QITE requires about 500 and 1000 updates, respectively (purple). }
    \label{fig:sup_Err}
\end{figure*}

As the first application, we implement QFM to simulate the time evolution of quantum states with nontrivial topology, which appears when the classical data set is encoded in quantum states~\cite{PhysRevLett.122.040504,lloyd2016quantum,huang2021power}. Specifically, we focus on the rotational dynamics of the $Z$-axis when encoding classical ring-shaped data distributions into quantum states, preserving their topological connectivity through rotations of $\mathcal{T}=20$ steps (  Fig.~\ref{fig:sim_ring}a). The evolution  starts from $S_{0}=\{\ket{\psi_{0}^{m}}\left|\ket{\psi_{0}^{m}}=e^{-i\sigma_{x}G_{m}}\ket{0}\right .,G_{m}\sim U(0,2\pi)\}$ to $S_{\tau}=\{\ket{\psi_{\tau}^{m}}\left|\ket{\psi_{\tau}^{m}}=e^{-i\sigma_{z}G_{\tau}}\ket{\psi_{\tau-1}^{m}}\right .,G_{\tau}=\pi \tau/\mathcal{T}\}$. The loss function of $l$-th step is chosen as:
\begin{equation}
    D(\vec{\theta}_{l})=1/M\sum_{m=1}^{M}\braket{\psi_{\tau}^{m}|V(\vec{\theta}_{\tau})|\widetilde{\psi}_{\tau-1}^{m}},
\end{equation}
leading to the generated ensembles in  Fig.~\ref{fig:sim_ring}a. The performance of QFM is validated by the match of the deviation $\braket{Y}^{2}$ from the theoretical and generated results (  Fig.~\ref{fig:sim_ring}b).

\subsection{2. Entanglement growth}

Through dynamically tuned multi-body interactions within a layered quantum-circuit architecture, QFM enables entanglement growth from separable to maximally entangled states, a key capability for scalable quantum-network node preparation and fault-tolerant logic-gate implementation~\cite{entangle_1,entangle_2,entangle_3,entangle_4}. We apply QFM to grow the entanglement entropy in $2$- and $3$-qubit quantum systems~\cite{ent_growth} over $10$ steps. The loss function at the $\tau$-th step is defined as
\begin{align}
D(\vec{\theta}_{\tau})=\frac{1}{M}\sum{m=1}^{M}\sum_{i=1}^{3}(e_{i}-\widetilde{e}^{m}_{i}(\vec{\theta}_{\tau}))^{2},
\end{align}
where $\widetilde{e}_{i}^{m}(\vec{\theta}_{\tau})$ denotes the entanglement entropy of the dynamically generated state $\ket{\widetilde{\psi}_{\tau}^{m}}=V(\vec{\theta}_{\tau}^{op})\ket{\widetilde{\psi}_{\tau-1}^{m}}$ in the ensemble $\widetilde{S}_{\tau}$, and the target value $e_{i}$ increases linearly as $\tau/\mathcal{T}$. As shown in  Fig.~\ref{fig:sim_entangled}b,c, the ensemble-averaged entanglement entropy closely follows the target linear growth. Figure~\ref{fig:sim_entangled}d shows the distribution of states with given entanglement entropy at each step. In terms of circuit depth, QFM maintains the same depth for both the $2$-qubit and $3$-qubit systems. Therefore, for larger systems, if the target entanglement can be decomposed into multiple two- or three-body correlations, the required circuit depth scales linearly with the number of entangled units. When such a decomposition is not possible and QFM is trained directly to increase the entanglement of the entire system, the circuit depth still increases moderately.

\subsection{3. Magnetic Phase Transition}
Magnetic phase transitions, driven by quantum-fluctuation-induced symmetry breaking, reveal universal scaling in collective spins and guide quantum material design with controlled quantum phases~\cite{TFIM_1,TFIM_2}. For a TFIM described by the Hamiltonian:
\begin{align}
    H(g_{\tau})=-\sum_{i}\sigma_{z}^{i}\sigma_{z}^{i+1}-g_{\tau}\sum_{i}\sigma_{x}^{i},
    \label{eq:TFIM}
\end{align}
which undergoes a phase transition from the ordered ferromagnetic phase to the disordered phase with $g=1$ (  Fig.~\ref{fig:sim_TFIM}a). We apply QFM to generate the ground state ensemble $\widetilde{S}_{\tau}=\{\ket{\widetilde{\psi}_{\tau}^{m}}\left|\ket{\widetilde{\psi}_{\tau}^{m}}=U(\vec{\theta}_{\tau}^{op})\ket{\widetilde{\psi}_{\tau-1}^{m}}\right .\}$ of $H(1.5\tau/\mathcal{T})$, where $\tau=1\to 15$ and $\mathcal{T}=15$. The loss function of step $\tau$ is: 
\begin{align}
    D(\vec{\theta}^{op}_{\tau})=&(1/M)\sum_{m=1}^{M}\bra{\widetilde{\psi}_{\tau-1}^{m}V^{\dagger}(\vec{\theta}_{\tau})}\notag\\
&H(1.5\tau/\mathcal{T})\ket{V(\vec{\theta}_{\tau})\widetilde{\psi}_{\tau-1}^{m}}.
\end{align}
In  Fig.~\ref{fig:sim_TFIM}b, we demonstrate the performance in a 4-qubit system, generating ground state ensembles with QFM and measuring magnetization $\mathcal{M}=\left(\sum_{i}\sigma^{i}_{z}\right)/n$. In contrast to QuDDPM, which only generates the state ensemble with specific magnetization, QFM can track varying average magnetization $\bar{\mathcal{M}}$ of state ensembles, as shown in  Fig.~\ref{fig:sim_TFIM}c with various qubits $n=2\to 11$ and field strengths $g_{\tau}=0.0\to 1.5$, which matches the theoretical result (  Fig.~\ref{fig:sim_TFIM}d).

\section{APPENDIX G: ERROR AND ADVANCEMENT IN LARGER SYSTEMS}

To ensure that QFM continues to generate high-fidelity results and maintains a reduced number of circuit adjustments for different system sizes, we performed free-energy estimation for systems of varying sizes. In both the four-qubit ( Fig.~\ref{fig:sup_Err}a) and six-qubit cases ( Fig.~\ref{fig:sup_Err}c ), QFM accurately reproduces the free-energy values while requiring only about $40\%$ of the circuit adjustments needed by conventional methods ( Fig.~\ref{fig:sup_Err}b and  Fig.~\ref{fig:sup_Err}d ). This confirms that the efficiency advantage of QFM is intrinsic to the method and scales favorably with system complexity, and that the error does not grow rapidly with system size.

%

\end{document}